\documentclass[reprint, superscriptaddress, amsmath, amssymb, aps, prm, floatfix, showkeys]{revtex4-2}
\usepackage{float}
\usepackage{graphicx}
\usepackage{dcolumn}
\usepackage{bm}
\usepackage{lipsum}
\usepackage{siunitx}
\usepackage{amssymb}
\usepackage{hyperref}
\hypersetup{colorlinks=true, linkcolor=blue, filecolor=blue, urlcolor=blue, citecolor=blue }
\usepackage{xcolor}
\usepackage{amsmath}
\usepackage{appendix}

\begin{document}

\title{Brittleness of metallic glasses dictated by their state at the fragile-to-strong transition temperature}

\author{Achraf Atila}
\email{achraf.atila@uni-saarland.de}
\author{Sergey V. Sukhomlinov}
\author{Marc J. Honecker}
\author{Martin H. Müser}
\affiliation{Department of Material Science and Engineering, Saarland University, Saarbr\"{u}cken, 66123, Germany}

\date{\today}

\begin{abstract} 
The effect of cooling on the brittleness of glasses in general, and bulk metallic glasses (BMGs) in particular, is usually studied with continuously varying cooling rates; slower cooling rates lead to stiffer, harder, and more brittle glasses than higher cooling rates. These protocols obscure any potential discontinuity that a glass might experience depending on whether its microstructure resembles that of a fragile or a strong glass-forming liquid. Here, we use large-scale molecular dynamics to simulate the nanoindentation behavior of model BMGs (Zr$_{0.6}$Cu$_{0.3}$Al$_{0.1}$) obtained by rapidly quenching equilibrium melts from temperatures above and below the fragile-to-strong transition temperature $T_\textrm{fst}$, leading to fragile and strong glasses, respectively. 
While the contact modulus deduced from the indentation simulation evolves smoothly with the temperature $T_\text{q}$ from which the melt is quenched, the plastic response changes quasi-discontinuously as $T_\text{q}$ passes through $T_\textrm{fst}$. In particular, strong glasses develop highly asymmetric flow profiles with mature shear bands, in contrast to fragile glasses. Quantitative differences reveal themselves not only through a formal von Mises localization parameter analysis but also through image analysis of flow patterns using pre-trained artificial intelligence models. Moreover, seemingly erratic flow profiles for our indentation geometry produced surprisingly reproducible and, thus, deterministic features.
It remains to be determined to what extent other classes of glass formers follow our observation that the degree of brittleness is significantly influenced by whether the melt is fragile or strong when it falls out of equilibrium at the glass transition temperature.
\end{abstract}


\maketitle

\section{\label{sec:intro}Introduction}
Bulk metallic glasses (BMGs) have elastic strain limits of up to $5\%$~\cite{Glushko2024NC} and high tensile~\cite{Inoue2003} as well as yield~\cite{Ritchie2011} strengths of order 1~GPa exceeding that of crystalline and poly-crystalline metals by an order of magnitude.
At the same time, BMGs have defied the rule that materials cannot be simultaneously strong and tough~\cite{Ritchie2011}.
In fact, BMGs are known to have the highest known damage tolerance~\cite{Ritchie2011}, which is the product of fracture yield strength and toughness. 
On the downside, BMGs generally strain to soften, making them prone to localized deformations in the form of shear bands~\cite{sopu2023JAC, Mu2021ADM, Liu2013Acta}, whereby undesirable surface markings can be produced, which ultimately evolve into cracks. This contrasts conventional metals, which usually strain hardens as the dislocations are the main carriers of plasticity~\cite{Xie2023SM, Xie2023PRM}. This behavior limits the moldability of most BMGs and their widespread technological applicability.

One subject of ongoing research is how the thermal history of a BMG affects its atomic structure and, thereby, its deformation mechanisms, most notably shear bands~\cite{sopu2023JAC, opu2020SM, Walley2007MMTA}. 
It is established both experimentally~\cite{Yang2022JNC} and from simulations~\cite{Shi2007acta,Zhang2022PNAS} that (bulk-metallic) glasses produced with a small cooling rate are stronger but also more prone to shear banding than those quenched more quickly. 
In contrast, silica-based oxide glasses are always brittle.
A distinguishing feature of silica-based glasses is the ratio of the fragile-to-strong transition (FST) temperature $T_\textrm{fst}$ and the glass transition temperature $T_\textrm{g}$, i.e., typical values of  $T_\textrm{fst}/T_\textrm{g}$ clearly exceed unity for silicates but are of order unity for BMGs. 
This prompts us to hypothesize that the brittleness of a glass is significantly influenced by the degree to which its atomic structure resembles that of a strong or fragile melt.
In particular, the question arises whether the quasi-discontinuity of both dynamic and static properties of the melt at the FST translates into discontinuities in the mechanical properties of the corresponding glass.

A few notes on the FST, which has been argued to occur in any glass-forming melt~\cite{Angell1991NOC}, and the glass transition itself may be in place. 
The (equilibrium) viscosity and atomic diffusion coefficients cross over from a non-Arrhenius to an Arrhenius dependence upon cooling in a rather narrow temperature window near $T_\textrm{fst}$, which is accompanied by a breakdown of the Stokes-Einstein relation below $T_\textrm{fst}$~\cite{Angell1995, Angell2000LAP}.
The specific heat of melts $c_p$ at $T>T_\textrm{fst}$ is substantially above that associated with the rule of Dulong-Petit.
This means that structural rearrangements go beyond (quasi-) harmonic vibrations so that the melt is called (thermodynamically) \textit{fragile}.
After passing through a maximum near $T_\text{fst}$ upon cooling, $c_p$ quickly decreases to values satisfying Dulong-Petit.
This is indicative of vibrations around barely changing reference structures.
The melt is called \textit{strong}.
The potentially most localized abrupt change at the FST is the jump of the correlation length describing the asymptotic decay of density oscillation~\cite{Stolpe2016PRB, Sukhomlinov2018PRM}.
In contrast to the FST, the glass transition refers to a more continuous process in which a melt falls out of equilibrium during cooling, however, with a strong dependence on the cooling rate.
Similar to the FST, the glass transition leads to a reduction of the specific heat toward the rule of Dulong-Petit. 

Ascertaining to what degree FST discontinuities translate into corresponding features in the glass requires \textit{equilibrium} melts to be quenched instantly from both above and below $T_\text{fst}$ to well below $T_\text{g}$ in the first step.
This way, the glass represents the frozen-in structure of a fragile and strong melt, respectively, and hence will be called either a fragile glass or a strong glass in the following. 
Whenever $T_\text{g}$ and $T_\text{fst}$ are not clearly separated, conventional cooling protocols using constant cooling rates plus potentially holding periods below $T_\text{g}$ generally produce a mix of fragile and strong glass, since the melt would not be fully equilibrated at $T > T_\text{fst}$, while some low-temperature relaxations would occur at $T < T_\text{fst}$.
The lower the cooling rate $\dot{q}$, the more the structure leans toward a strong glass, but changes are continuous in $\dot{q}$.
Thus, previous studies~\cite{Zhang2022PNAS, Ghaemi2022JNCS, JafaryZadeh2017} revealing that the failure mechanisms evolves smoothly from being homogeneous in a high-$\dot{q}$ sample to localized shear bands using low $\dot{q}$ do not contradict our hypothesis that the brittleness of a (bulk metallic) glass is dictated by its state at the FST.

In the remainder of this work, we use molecular dynamics (MD) simulations to contrast the mechanical behavior of strong and fragile glasses during nanoindentation.
To this end we recycle previously produced atomistic equilibrium configurations of a (\textit{in-silico} model for a) ternary BMG, Zr$_{0.6}$Cu$_{0.3}$Al$_{0.1}$, which had been shown to undergo a FST in thermal equilibrium~\cite{Sukhomlinov2018PRM}. 
MD is an ideal tool for this study because equilibrium melts can be quenched extremely quickly and homogeneously, preventing temperature-gradient-induced heterogeneities or residual stresses from occurring.

\section{\label{sec:method}Methods}

\subsection{Molecular dynamics}
In this work, we simulate the ternary alloy Zr$_{0.6}$Cu$_{0.3}$Al$_{0.1}$ using an embedded-atom potential~\cite{Daw1984PRB} parameterized by Cheng, Ma, and Sheng~\cite{Cheng2009PRL}.
This system revealed a transition from a fragile to a strong liquid at $T_\text{fst} \approx \SI{830}{\kelvin}$~\cite{Sukhomlinov2018PRM}, at which the correlation length of the long-range density fluctuations changed quasi-discontinuously by roughly 10\% in a very narrow temperature range similarly as in a related experimental BMG called Vitleroy 106a~\cite{Stolpe2016PRB}.

Initial configurations were taken from the above-mentioned previous work~\cite{Sukhomlinov2018PRM}, in which a sample containing $\approx$ 8800 atoms was equilibrated at various temperatures, including a narrow temperature regime below $T_\text{fst}$.
The full equilibration times used during step-wise cooling were:
30~ns at $T = 1,000$~K, 64~ns at 900~K, 400~ns at 850~K, 1.25~$\mu$s at 812~K, 2.5~$\mu$s at 800~K, and 4.0~$\mu$s at 787~K. 
In the following, samples obtained by quenching an equilibrium sample from $T < T_\text{fst}$ will be referred to as \emph{strong} glasses, while those obtained from equilibrium simulations at $T>T_\text{fst}$ will be called \emph{fragile} glasses.
All samples were duplicated $n_x=3, n_y=3, n_z=3$ times in $x$, $y$, and $z$-direction, respectively, equilibrated for $\SI{4}{\nano\second}$ at the temperature from which they were obtained, and subsequently quenched to room temperature (T = $\SI{300}{\kelvin}$ with cooling rate of $\SI{1}{\kelvin/\pico\second}$, and annealed for another $\SI{4}{\nano\second}$. During these simulations, a timestep of $\SI{2}{\femto\second}$ was used with periodic boundary conditions applied to all box directions.
Such equilibrated samples were replicated again to obtain a larger sample with 230 x 16.5 x 86 nm$^3$, used in the nanoindentation simulations. Using these large samples ($\approx$ 18 Mio atoms), a surface was created along the $z$-axis, and a 1~nm thick layer of atoms at the bottom of the simulation box along the z-direction was fixed. The samples were periodic along the $x$- and $y$-direction. Next, the samples were further equilibrated for $\SI{200}{\pico\second}$ at room temperature and 0 MPa using a timestep of $\SI{2}{\femto\second}$. The temperature was maintained using a Langevin thermostat~\cite{Schneider1978PRB} while the pressure was controlled using Nosé-Hoover chain barostat~\cite{Shinoda2004PRB}. All simulations were conducted using the LAMMPS simulation package~\cite{Thompson2022CPC} and the snapshots were prepared using OVITO~\cite{Stukowski2009MSMSE}.

The indentation was performed with a rigid cylindrical indenter exerting a force of magnitude $F(r) = -K(r - R)^2$ on each atom, where $K$ is the force constant which we chose to be $K$ = 10 eV/\text{\AA}$^3$, $r$ is the distance from the center of the indenter to the respective atom, and $R$ is the radius of the indenter which was fixed to $R$ = $\SI{100}{\nano\meter}$. Lilleodden \textit{et al.}~\cite{Lilleodden2003LMPS} studied the response of single-crystalline gold to nanoindentation while varying the force constant of the indenter and found that the force-displacement curves are unaffected by the change of $K$.
Displacement-controlled nanoindentation simulations were performed by displacing the indenter along the $z$-direction towards the sample with a fixed velocity of $v$\textsubscript{i} = $\SI{5}{\meter/\second}$ until reaching a depth of $\SI{20}{\nano\meter}$ and using a timestep of $\SI{2}{\femto\second}$.

\subsection{Local strain analysis}
The detection of shear bands was performed by using the von Mises shear strain. The von Mises strain was calculated by getting an affine transformation matrix to the relative neighbors through a least-square fitting~\cite{Shimizu2007MTJ}. The undeformed glass was chosen as a reference configuration. More details of the strain calculation can be found in Refs.~\cite{Stukowski2009MSMSE, Shimizu2007MTJ}. 
To get the regions with a nonaffine displacement, the current atomic configuration needs to be compared with a reference configuration. The neighbors of each particle are found by defining a sampling radius (i.e., 10~\text{\AA} in our case). The local strain is then calculated from the displacements of the neighboring particles to the central one and the relative displacements that they would have if they were in a region of uniform strain. The final value of the nonaffine displacement was divided by the number of atoms within that spherical region of the same radius. More details about the calculation of the nonaffine displacements are given in Ref.~\cite{Falk1998}.

\subsection{Image analysis}

Snapshots of the local von Mises shear strain taken at different depths during the nanoindentation process were image analyzed.
To this end, pre-trained DINOv2 neural networks~\cite{oquab2023dinov2} were used for feature extraction because DINOv2 is known for its robustness in capturing image features through self-supervised learning.
The DINOv2 architecture is based on the so-called vision transformer (ViT)~\cite{dosovitskiy2021imageworth16x16words}, which is currently the \textit{de-facto} standard for computer vision.
In this work, we mainly use the bare backbones ViT-b/14 with registers~\cite{darcet2024visiontransformersneedregisters} and ViT-g/14 without registers available at GitHub (\url{https://github.com/facebookresearch/dinov2}), because they showed the best results for our analysis.
The models had been pre-trained on 142 million images in a contrastive learning setting, which allows the model to capture robust image features, although images supposedly did not show shear bands or flow lines. 

Each image was preprocessed to meet the input requirements of the DINOv2 model.
The preprocessing steps included resizing the image to 224$\times$224 pixels, converting it to a PyTorch~\cite{paszke2019pytorchimperativestylehighperformance} tensor, and applying the DINOv2-specific preprocessing function.
The latter includes normalizing the images by the mean and standard deviation values of ImageNet~\cite{russakovsky2015imagenetlargescalevisual} statistics: mean = [0.485, 0.456, 0.406] and standard deviation = [0.229, 0.224, 0.225].
After putting all images through the model, the cosine similarity of each pair of images was computed.
It ranges from -1 (anti-parallel, minimum similarity) to  1 (parallel, maximum similarity).

\section{\label{sec:results}Results}

In this work, we simulate the indentation process of glasses obtained from quenching equilibrium melts from six temperatures $T_q$ to $T = 300$~K, where all indentation simulations were conducted.
Three of the initial temperatures, that is, $T_1 = 787$~K, $T_2 = 800$~K, and $T_3 = 812$~K were below the fragile-to-strong cross-over temperature $T_\text{fst} = 830$~K, the remaining ones $T_4 = 850$~K, $T_5 = 900$~K, and $T_6 = 1,000$~K were above  $T_\text{fst}$.
To be concise, the sample quenched from $T_1$ may not have fully reached equilibrium at $T_1$.
To be specific, the relaxation time of energy fluctuations in the $\alpha$-relaxation regime can be estimated to be $\tau = 10$~ns and the Kohlrausch exponent to $\beta = 0.73$, which yields a $1-\exp(-(t_\textrm{sim}/\tau)^\beta) = 0.995$ complete $\alpha$ relaxation given an simulation time of $t_\textrm{sim} = 100$~ns.
For the volume calculation, the relaxation is even more complete.
In this case, $\tau = 1$~ns and $\beta = 0.79$ yielding $1-\exp(-(t_\textrm{sim}/\tau)^\beta) = 1$.

\begin{figure}[h]
\centering
\includegraphics[width=\columnwidth]{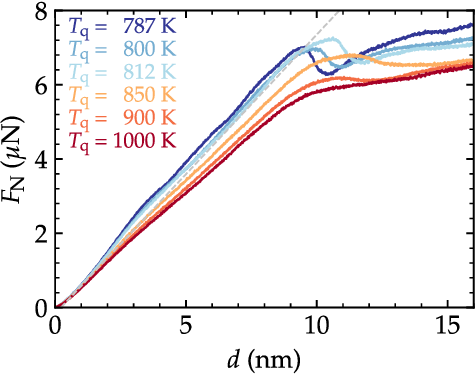}
\caption{Force displacement curves of loading for the fragile glasses and the strong glass at $T=\SI{300}{\kelvin}$. The dashed line reflects an indentation with a contact modulus of $E^* = 57.9$~GPa. Deviations of the data from linearity at very small $d$ are due to an effectively finite-range repulsion between the indenter and samples.}
\label{fig:forceDisplacementGeneric}
\end{figure}

At first glance, all six samples show qualitatively similar force-displacement curves  as revealed in Fig.~\ref{fig:forceDisplacementGeneric}:
an initial Hertzian regime up to an indentation of $d \approx 2$~nm, followed by a non-elastic regime.
A maximum indentation force of $F \approx 7 \pm 0.5~\mu\text{N}$ is reached at $d \approx 10 \pm 1$~nm for all  $T_\text{q} \le 900$~K samples. 
After a short drop, $F$ starts increasing again at $d \approx 12 \pm 2$~nm. 

Closer inspection of Fig.~\ref{fig:forceDisplacementGeneric} reveals qualitative differences between fragile and strong glasses.
All strong samples have similar $F_\textrm{N}(d)$ relations at $d < d_\text{max}$, where $d_\text{max}$ is the depth at which the force is a local maximum.
In contrast, $F_\textrm{N}(d)$ of fragile samples reveal some $T_\text{q}$ dependence at $d < d_\textrm{max}$.
The situation reverts when $d$ exceeds $d_\textrm{max}$ by a few nanometers.
This time, $F_\textrm{N}(d)$ converges to a single dependence for all fragile glasses but reveals a weak $T_\text{q}$ dependence for the strong glasses.
Moreover, the drop in the indentation force after reaching its maximum is noticeably more pronounced for strong than for fragile glasses. 

\begin{figure}[hbt!]
\centering
\includegraphics[width=0.9\columnwidth]{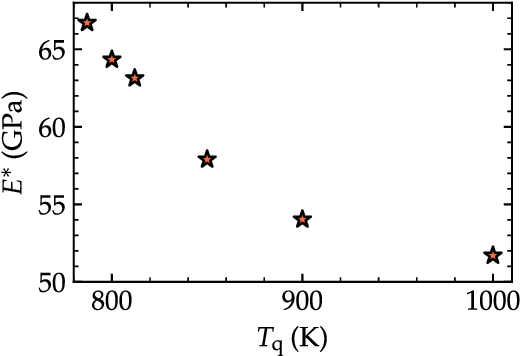}
\caption{Contact modulus $E^*$ at $T = 300$~K as a function of $T_\text{q}$ as deduced from the initial ($0.2 < d/\text{nm} < 2.5$) linear $F_\textrm{N}(d)$ dependence of the indentation process using $F_\textrm{N}(d) \approx \frac{\pi}{4}E^* Ld$. The very initial  $d < 0.2$~nm domain is excluded from fit because the finite stiffness of the cylinder-BMG repulsion causes a parabolic dependence at these extremely small indentation depths.
}
\label{fig:Estar}
\end{figure}

Subtle qualitative differences between the fragile and the strong glasses can already be noticed at a relatively small indentation depth.
The strong glass starts to deviate from their linear response by assuming a left-curved $F_\text{N}(d)$ relation, while the fragile glasses adopt a minor right-curved $F_\text{N}(d)$ dependence. 
In contrast, the contact modulus $E^*(T_\text{q})$ has a smooth dependence on $T_\text{q}$ without detectable signs of a discontinuity of the function itself or its derivative, as is revealed in Fig.~\ref{fig:Estar}
These findings align with the existing literature, which also reports more significant effects of cooling rate on plastic than on elastic properties~\cite{Zhang2022PNAS}.

\begin{figure*}[hbt!]
\centering
\includegraphics[width=\textwidth]{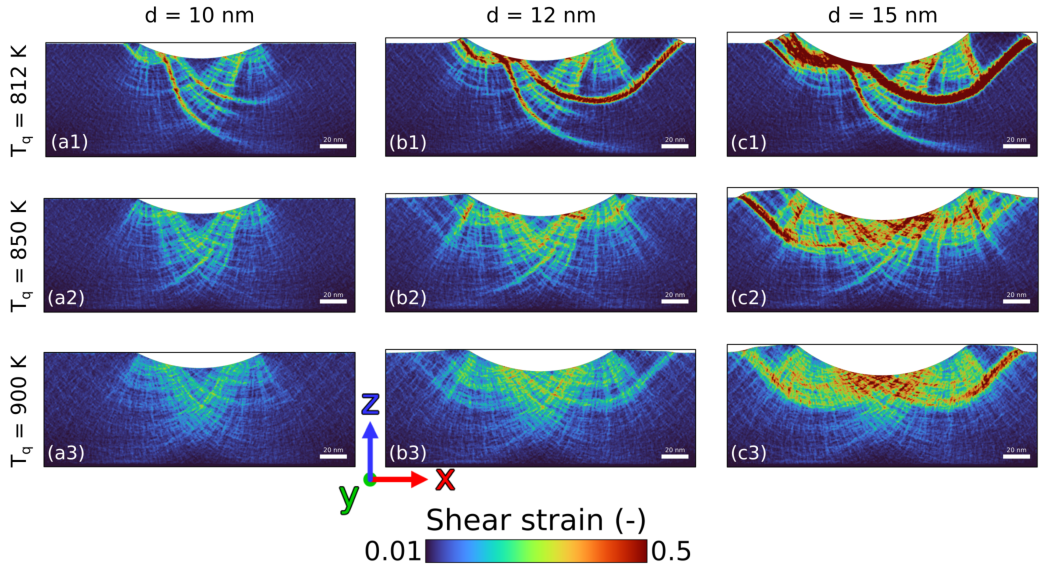}
\caption{Local shear strain maps for the strong and fragile glasses loaded to an indentation depth of (a) $d$ = 10~nm, (b) $d$ = 12~nm, and (c) $d$ = 15~nm.}
\label{fig:ShearStrainMap}
\end{figure*}

Despite subtle differences between fragile and strong plasticity at the early loading stages, effects are more pronounced at large indentation.
To analyze plasticity, we investigated the local von Mises local shear strain, as described in Sec.~\ref{sec:method}.
Exemplary results are shown in Fig.~\ref{fig:ShearStrainMap} with representatives corresponding to the three different plasticity regimes of the $F(d)$ curve, i.e., before, near, and after the maximum indentation force.
Fig.~\ref{fig:SnapshotsVMSSI} depicts more snapshots of the von Mises strain, i.e., for all six samples at seven depths, including one snapshot taken in the elastic regime. 
We also note that visualization of the shear bands using non-affine displacement squared, defined similarly as by Falk and Langer~\cite{Falk1998} but normalized to the number of neighbors, produced snapshots, which closely resembled those based on the von Mises strain, as is evidenced in Fig.~\ref{fig:SnapshotsDminSI}.

Shear bands, which are the main carrier of plasticity in metallic glasses~\cite{Hassani2019PRL, Liu2013Acta, Greer2013MSER, Yang2016Acta} start to occur once $F(d)$ deviates from the elastic response.
They mature, i.e., the shear bands reach a stage where their growth slows down or stops after $F(d)$ reaches its maximum, at which point they have hit the surface.  
Contrary to previous simulations~\cite{Yang2016Acta,Adjaoud2021, Zhao2021CMS, Zhang2022PNAS, sopu2023JAC} and our fragile glasses, the \textit{strong} samples develop visible, radially shaped shear bands with an average width of $w \approx$ 8~nm.
This is on par with $w \approx$10 nm~\cite{Sheng2022PRL, Greer2013MSER, Zhang2006APL},  which is the intrinsic thickness of mature shear bands in \textit{ex-silico} metallic glasses.
The shear bands of strong glasses are more localized and more asymmetric than in fragile glasses, to the extent that they could be said to differ qualitatively. 

\begin{figure}[hbtp]
\centering
\includegraphics[width=0.9\columnwidth]{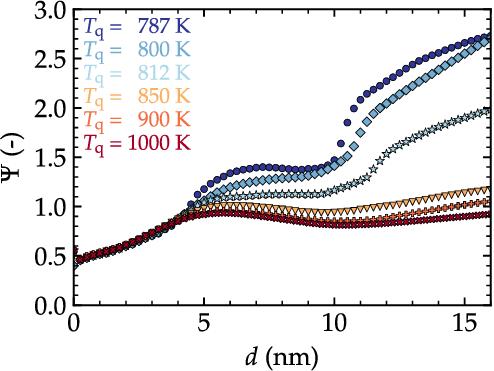}
\caption{Strain localization parameter $\Psi$ as a function of the indentation depth for the strong and fragile glasses. Different symbols are used for their respective temperatures $T_\text{q}$: circles for 787~K, diamonds for 800~K, stars for 812~K, triangles down for 850~K, pluses for 900~K, and crosses for 1000~K.}
\label{fig:Strainlocann}
\end{figure}

Going beyond a visual inspection of the shear bands requires a quantitative analysis to be conducted. 
To this end, we study the standard deviation of the atomic von Mises strain divided by the average atomic von Mises strain,  $\Psi = \langle \Delta \eta_n^2 \rangle^{1/2} / \langle \eta_n \rangle$, where $\eta_n$ is the von Mises strain projected to atom $n$,   $\langle ...\rangle$ an average over all atoms, and $\Delta \eta_n = \eta_n - \langle \eta_n \rangle$.
$\Psi$ is similar to a previously defined localization parameter~\cite{Cheng2009Acta} but normalized so that it does not change when atoms with zero von Mises strain are added. 
Like the previous parameter, $\Psi$ is sensitive to localization because atoms with large variance tend to cluster in shear bands. 
Results for $\Psi$ are shown in Fig.~\ref{fig:Strainlocann}.
At very small $d$, the value for $\Psi$ is essentially identical for all samples.
This can be understood from the fact that elastic deformation is scale-invariant and, thereby, the value for $\Psi$ at $d = 0.5$~nm is merely 2\% 
between the most fragile and the strongest sample compared to a 25\% difference in $E^*$.
 
The values of $\Psi(d)$ bifurcate at a depth of $d \gtrsim 4$~nm, where $\Psi(d)$ keeps increasing for the strong samples while they saturate otherwise.
An even stronger qualitative difference of strong and fragile samples becomes apparent at $d \gtrsim 9$~nm, where the slope of $\Psi(d)$ suddenly increases near the point of the local $F_N(d)$ minima for the strong samples, while $\Psi(d)$ has no such discontinuity in the fragile samples.

From the results shown so far, a qualitative difference in the deformation behavior of glasses reflecting the structure of a fragile and a strong melt has become obvious.
To ensure that these differences are not merely a kinetic effect resulting from the ratio of the indentation rate and the relaxation time at $T_\text{q}$ surpassing some critical value, we performed another indentation simulation of the $T_\text{q} = 850$~K sample, however, using a ten-fold increased indentation velocity.
This exceeds the ratio of relaxations times between the ``hottest strong'' and the ``coldest fragile'' melt, for both energy and volume relaxation. 

\begin{figure}[h]
\centering
\includegraphics[width=0.95\columnwidth]{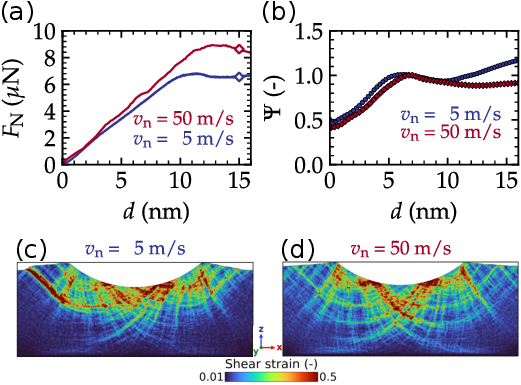}
\caption{Contrasting the indentation process of the $T_\text{q} = 850$~K  fragile glass using two different indentation velocities. Panel (a) shows the normal force $F_\text{N}$ and (b) the strain-localization parameter $\Psi$ as a function of displacement $d$ for both velocities. 
Panel (c) and (d) depict the von Mises shear strain for $v_0 = 5$~m/s and $v_0 = 50$~m/s, respectively.}
\label{fig:Load_Displacement_50_5mps}
\end{figure}

The ten-fold increased indentation velocity produced a noticeable increase in the $F_\text{N}(d)$ relationship.
However, the higher indentation velocity produces noticeably more symmetric and less localized shear bands or rather flow patterns, as is revealed in Fig.~\ref{fig:Load_Displacement_50_5mps} both visually but also in terms of the strain-localization parameter.
The higher symmetry of the flow images at the higher velocity could be argued to arise from the reduced time for an initial local symmetry breaking to migrate from small scales to large scales.
In our simulations, the symmetry breaking arises due to structural heterogeneities, potentially also, to some degree, by thermal fluctuations.
In case of instability, which is unavoidable in strain-softening materials, numerical round-off errors, even in a perfect symmetric finite-element simulation, might suffice to cause similar asymmetries in the shear bands as in the MD simulations when the loading is slow.

\begin{figure*}[]
\centering
\includegraphics[width=1.0\textwidth]{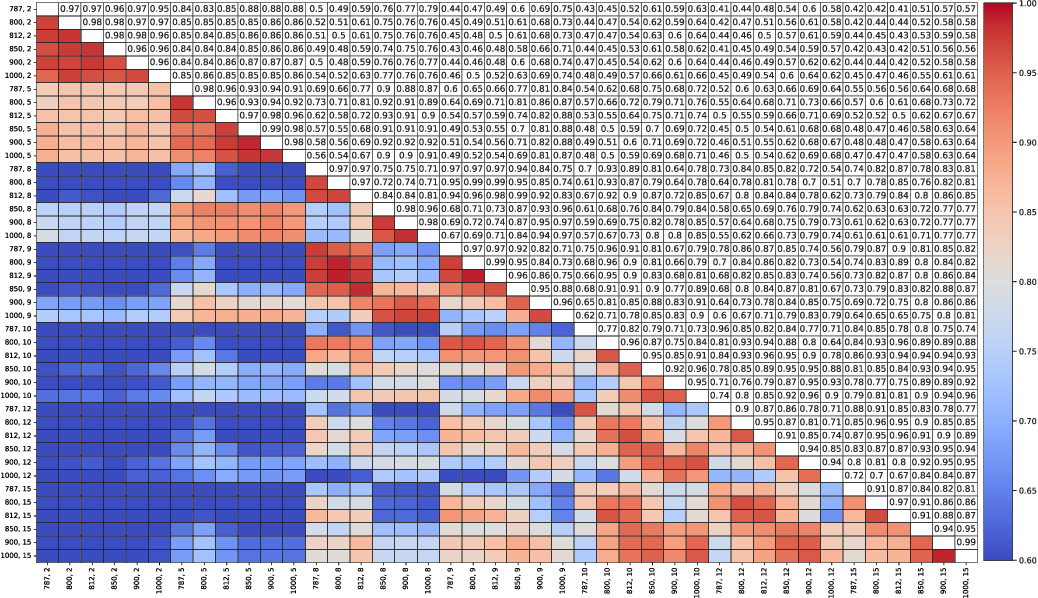}
\caption{Cosine similarity matrix using ViTg14 for the images shown in Fig.~S1. The first number in each row and column gives $T_\text{q}$ in Kelvin, while the second gives $d$ in nm.}
\label{fig:Mat}
\end{figure*}

To explore the degree to which our findings can be tested experimentally if images of flow profiles are available but there is no spatially resolved von Mises strain, we quantified the similarity between the shear-band signature using two different models, as described in the Methods section. 
The result produced with ViT\_g/14 is shown in Fig.~\ref{fig:Mat}.
At each indentation depth, inspection of both colors and numbers included in the figure generally gives the largest difference in similarity between the hottest strong and the coolest fragile samples. 
This claim holds for both models.
Moreover, the ViT\_g/14 judges the similarity between the upper left and upper right panel in Fig.~\ref{fig:ShearStrainMap} to be more than that between the top right and the bottom right, i.e., 0.94 and 0.88 cosine similarity, respectively. 
This is an interesting finding given that the covered colored area is similarly large for the two samples indented to 15~nm.

Figuratively speaking, the human eye can clearly identify the top left panel as the child and the upper right as its adult, rugose version, while the upper and lower right panels represent two different adults. 

\begin{figure}[]
\centering
\includegraphics[width=0.67\columnwidth]{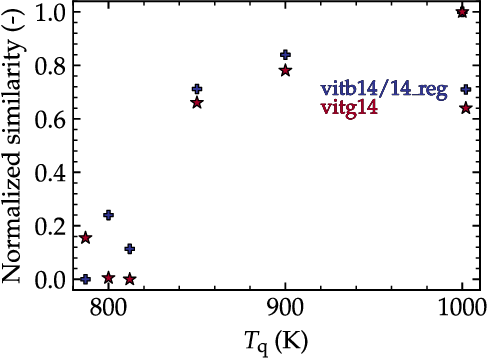}
\caption{Normalized similarities measure of a flow profile obtained from a glass quenched from various $T_\text{q}$ and from $1\,000$~K at $d = 15$~nm for the ViTb/14\_reg (plus symbols) and the ViTg/14 (star symbols) models. 
}
\label{fig:NormCorr}
\end{figure}

Since neither the color maps nor the numbers included in Fig.~\ref{fig:Mat} allow us to conclude whether the two models reveal discontinuities between fragile and strong glasses, a normalized similarity $S_\text{n}(T_\text{q})$ of a sample quenched from arbitrary $T_\text{q}$ and indented to $d = 15$~nm with that quenched from $T_\text{q} = 1,000$~K is shown in Fig.~\ref{fig:NormCorr}.
It is defined as the normalized cosine similarity between the sample quenched from $T_\text{q}$ = 1000~K and all other samples at an indentation depth of 15~nm, as defined by
\begin{equation}
\label{eq:similarity_rel}
S_\text{n}(T) = \frac{S^{1000\text{K}}_\text{cos}(T)- \min(S^{1000\text{K}}_\text{cos}(T_\textrm{q}))}{1- \min(S^{1000\text{K}}_\text{cos}(T_\textrm{q}))},
\end{equation}
where $S_\text{cos}^{T_1}(T_2)$ is the cosine similarity of the flow profiles of quench temperature $T_1$ and $T_2$.
This measure makes the two studied models operate on a similar domain, because it is normalized such that the similarity between the least similar and the $T_\text{q} = 1000$~K sample is zero.

Although the similarity indices show a clustering for the strong samples and a jump between strong and fragile samples, the judgment appears less clear than that by humans or the von Mises localization parameter. 
It is nevertheless remarkable that both models produce similar results,  although the ViT-b/14 has less than $8\%$ of the parameters compared to ViT-g/14.

\begin{figure*}[]
\centering
\includegraphics[width=\textwidth]{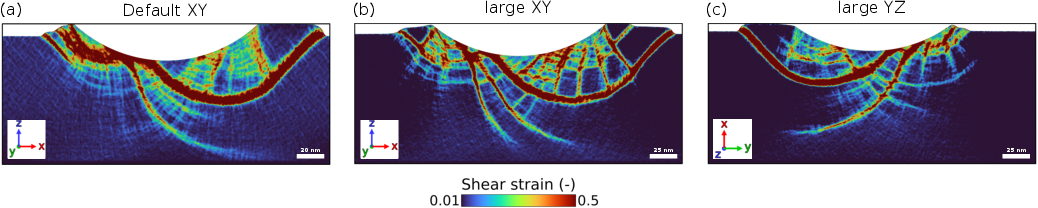}
\caption{Von Mises strain of (a) the default set up for $T_\textrm{q} = 812$~K and a depth of $d = 15$~nm and (b) and (c) for a slightly larger sample at a depth of  $d = \textcolor{blue}{??}$~nm. The starting configurations in (b) and (c) differed, i.e., the original glass sample in (c) was rotated by 90$^\circ$ before duplication so that the original $yz$ surface rather than the $xy$ surface was indented.  
}
\label{fig:diffsurf}
\end{figure*}

A final question to be addressed here is to what extent the shape of the flow profiles and shear bands, particularly those obtained for the strong glasses, is erratic or deterministic. 
All strong samples develop a long shear band to the right with spider-web-like features between the shear band and the surface, plus a shorter but thicker shear band to the left, which is very close to the surface.
The probability for breaking the symmetry when quenching from three independent samples each time in the same direction is 1/4, which is small compared to unity but not very small so that it is unclear whether some hidden memory in the strong melt causes the symmetry to be broken each time in the same direction. 
We remind the reader that all final configurations share the same ancestor, e.g., the sample obtained after equilibrating the melt at $T = 812$~K was quenched to yield the $T_\text{q} = 812$~K glass at 300~K, but also copied and further equilibrated at 800~K for times much exceeding the energy and volume relaxation time. 

Since equilibration and indentation simulations require several million CPU hours, averaging over independently produced samples is impractical.
To nevertheless ascertain how deterministic features are, we repeated the simulations for a larger cell, in which the quenched samples were not only repeated $14 \times 1 \times 5$ times in $x$, $y$, and $z$-direction respectively but $15 \times 1 \times 6$ times ($\approx$ 21.3 Mio).
Moreover, the original simulation cell kept its orientation one time but was rotated by 90$^\circ$ so that the original $yz$ surfaces became the new to-be-indented $xy$ surface.
Von Mises strains of these three different $T_\text{q} = 812$~K samples are contrasted in Fig.~\ref{fig:diffsurf}.
The two new configurations show the same characteristics as the old one, i.e., one prominent long shear band, which breaks the symmetry and has the spider-web feature underneath it. 
Thus, not only the gross features of the shear profile appear to be predetermined.

\section{\label{sec:discussion}Discussion}
The FST is conjectured to exist in all liquids~\cite{Ito1999Nature, Lucas2019NOCX}. MGs prepared experimentally are quenched with cooling rates that are fast enough to avoid crystallization and, at the same time, slow enough to allow for this FTS transition~\cite{Way2007Acta}. However, the glasses prepared in MD simulations are usually cooled with cooling rates several decades faster than those used in experiments~\cite{Atila2020me, JafaryZadeh2017, Atila2019PRB, Zhang2022PNAS, Atila2022PRB}, preventing them from experiencing the FTS liquid transition~\cite{Sukhomlinov2018PRM}. 
The results of this work show that this renders a one-to-one comparison between simulation and experiments problematic. 
However, suppose glasses are prepared in a way that makes them undergo FTS liquid transition. In that case, the differences between the simulated and experimentally prepared glasses will reduce, and a one-to-one comparison of the mechanisms will be much trusted. 
This was shown to be the case recently when authors combined MD and Monte Carlo swaps to equilibrate the system and produce stable glasses with properties closer to the experimental ones~\cite{Zhang2022PNAS}.

Regarding the mechanical behavior of MGs, Packard and Schuh~\cite{Packard2007acta} showed that for isotropic materials such as metallic glasses, the formation of narrow shear bands in the area with concentrated shear stresses that exceed the yield stress. Nevertheless, from our results presented above, the first shear transformation zones in the bulk sample appeared at an indentation depth of about 2.5~nm and $\approx$ 15 nm under the indenter, but not all of them are exactly along the axis of indentation, suggesting that the local shear stresses do not uniquely determine the STZ formation. Similar observations were made by Wang~\textit{et al.}~\cite{Wang2010CMS}. This can be explained by the fact that the local variation of the atomic configuration results in BMG having different zones with low or high shear stresses distributed in the sample and with STZ preferring to nucleate at the sites with low shear resistance~\cite{sopu2023JAC, Kosiba2020PRB}, which dictate the location where STZ form and also the propagation of the shear bands afterward. This is in line with the observations we presented, where asymmetric shear bands are observed in the strong glass, although the cylindrical indenter is symmetric. In addition to the strain field applied by the indenter, the shear band will naturally propagate towards the surface because it provides no constraints on their propagation~\cite{Wang2010CMS}.\\
The differences between strong and fragile glasses can be explained by the degree of stability of the glasses and the concept of the potential energy landscape (PEL)~\cite{Tang2021MH}. The strong glass was cooled using a much lower effective cooling rate $\approx 10^7$~K/s, giving the atoms more time to relax and move towards deeper energy minima. On the other hand, the fragile glasses were cooled from temperatures above the FTS transition temperature, where the samples did not explore deeper energy minima. Thus, those glasses are less stable than the strong ones. This implies that, when comparing the mechanical properties and the mechanical behavior of the strong and the fragile glasses, the strong glass will have larger mechanical properties and be more brittle, which is consistent with experimental and other simulation results that show that well-relaxed glasses have a rougher PEL, which results in lower atomic mobility, that makes the glasses more brittle and has a higher strength than the less stable glasses~\cite{Zhang2022PNAS, Tang2021MH}. 
Although their focus was not related to the FST, experimentally, it has been proven that the processing history of the glass dictates the yielding and the shear band dynamics in the sample~\cite{Liu2013ActaMat}. Shi and Falk~\cite{Shi2007acta,Shi2007TSF} made closely similar conclusions using binary Lennard-Jones glasses, where they observed a stronger localization of the local shear strain in the samples cooled slowly. \\
The presence of $\approx$ 8 nm thick shear bands in the strong glass, which are well compared to those found experimentally~\cite{Zhang2006APL}, is reported here for the first time. The radial shape of the shear bands is consistent with the slip-lines that develop during an indentation using a cylindrical indenter and are also observed experimentally~\cite{Antoniou2007JMR, Antoniou2005MSEA, SU2006Acta}. Previous MD simulations~\cite{Yang2016Acta, Zhao2021CMS} showed the presence of radial shear bands as well, but with a much lower thickness. These interesting differences in the deformation behavior of fragile and strong glasses shed light on the overall deformation mechanisms of MGs, where both localized and homogeneous deformation were observed for the same glass composition.

\section{\label{sec:conclusions}Conclusions}
In this paper, we used large-scale molecular dynamics simulations to explore how the fragility of the melt affects the brittleness of metallic glasses. We showed that melts quenched from just below and just above the fragile-to-strong transition temperature show qualitatively different plasticity. 
The strong glasses exhibit asymmetric and localized shear bands. In contrast, the fragile glasses form broader and less defined shear bands. These differences between the strong and the fragile glasses were quantitatively observed using a modified strain localization parameter as well as by using pre-trained artificial intelligence models for image similarities. The distinctness between the deformation behavior and plastic flow profile of the fragile and strong glasses arises because strong glasses can relax into deeper energy minima, resulting in more stable atomic configurations enhancing their mechanical strength.

These findings highlight a potentially overlooked link between the thermodynamic state of the liquid at the FST and the deformation behavior of the resulting glass. 
This link can be expected to generalize to other classes of glass-forming liquids than BMGs.
In fact, recent work suggested that the propensity for ductility is controlled by the topography of the energy landscape based on the simulation of various glass-forming systems~\cite{Tang2021MH}.
It changes from being relatively smooth above $T_\text{fst}$ (high $c_p$) to deep quasi-harmonic minima below $T_\text{fst}$ ($c_p$ in accordance with Dulong-Petit).
However, it remains to be determined to what extent the brittleness of other glass-forming systems changes abruptly depending on whether they are cooled from above or below $T_\text{fst}$.

\begin{acknowledgments}
The authors gratefully acknowledge the Gauss Centre for Supercomputing e.V. (www.gauss-centre.eu) for funding this project by providing computing time through the John von Neumann Institute for Computing (NIC) on the GCS Supercomputer JUWELS at Jülich Supercomputing Centre (JSC).
\end{acknowledgments}

\clearpage

\onecolumngrid
\begin{center}
\Large \textbf{Supplementary material to:\\Brittleness of metallic glasses dictated by their state at the fragile-to-strong transition temperature}
\end{center}

\renewcommand{\theequation}{S\arabic{equation}}
\renewcommand{\thefigure}{S\arabic{figure}}
\renewcommand{\thetable}{S\arabic{table}}
\setcounter{figure}{0} 

\begin{figure*}[hbtp]
\centering
\includegraphics[width=\textwidth]{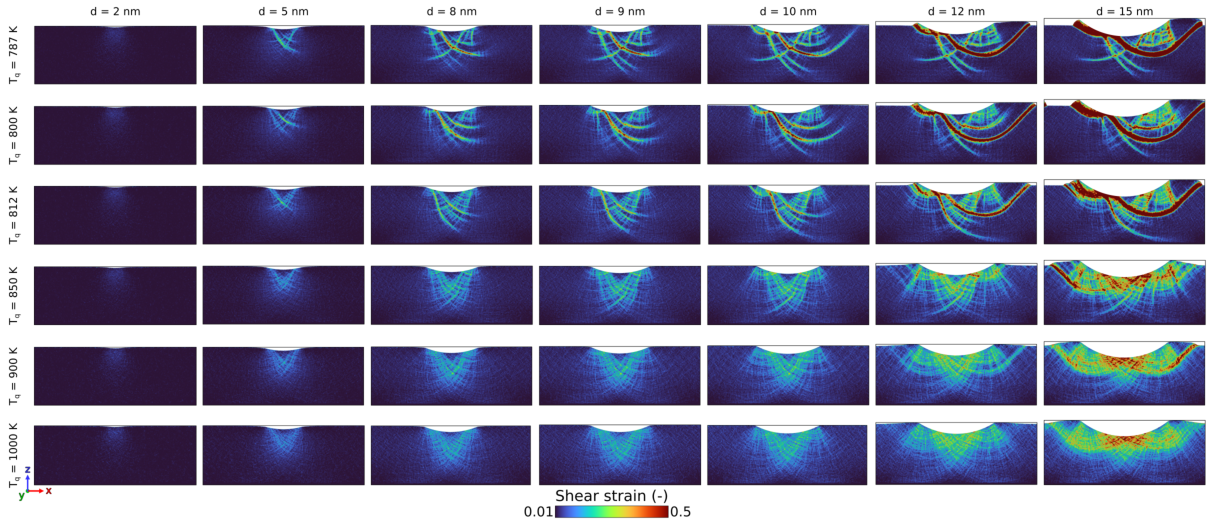}
\caption{Local shear strain maps for the strong and fragile glasses loaded to different indentation depths.}
\label{fig:SnapshotsVMSSI}
\end{figure*}

\begin{figure*}[hbtp]
\centering
\includegraphics[width=\textwidth]{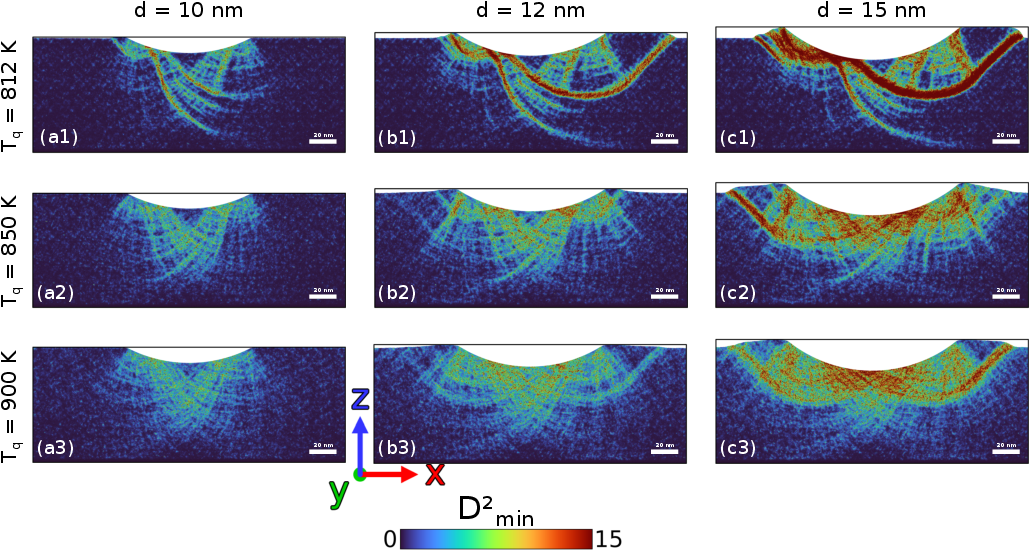}
\caption{Local non-affine displacement squared calculated as detailed in the main text for the strong and fragile glasses loaded to an indentation depth of (a) $d$ = 10~nm, (b) $d$ = 12~nm, and (c) $d$ = 15~nm.}
\label{fig:SnapshotsDminSI}
\end{figure*}

\end{document}